\documentclass[twocolumn,showpacs,amsmath,amssymb,prd]{revtex4}
             
\usepackage{graphicx}% Include figure files
\usepackage{dcolumn}% Align table columns on decimal point  
\usepackage{bm}% bold math
                
\def\beq{\begin{equation}}
\def\enq{\end{equation}}  
\def\ba{\begin{eqnarray}}
\def\ea{\end{eqnarray}}
\def\<{\langle}
\def\>{\rangle}

\begin{document}
\input{epsf}

\title{Neutrino mass hierarchy extraction using atmospheric neutrinos
  in ice} \author{Olga Mena$^{1,2}$, Irina Mocioiu$^3$ and Soebur
  Razzaque$^{4}$} \affiliation{$^1$ INFN Sez.\ di Roma, Dipartimento
  di Fisica, Universit\`{a} di Roma``La Sapienza'', P.le A.~Moro, 5,
  I-00185 Roma,Italy} \affiliation{$^2$ Institute of Space
  Sciences(IEEC-CSIC), Fac. Ciencies, Campus UAB, Bellaterra, Spain}
\affiliation{$^3$Department of Physics, Pennsylvania State University,
  University Park, PA 16802, USA} \affiliation{$^4$Space Science
  Division, Code 7653, U.S. Naval Research Laboratory, Washington DC
  20375, USA}

\begin{abstract}
We show that the measurements of 10 GeV atmospheric neutrinos by an
upcoming array of densely-packed phototubes buried deep inside the
IceCube detector at the South Pole can be used to determine the
neutrino mass hierarchy for values of $\sin^22\theta_{13}$ close to
the present bound, if the hierarchy is normal. These results are
obtained for an exposure of 100 Mton years and systematic
uncertainties up to 10\%.
\end{abstract}


\date{\today}
\maketitle

\section{Introduction}

Neutrino physics has undergone a true revolution over the last decade,
with a large number of experiments of different types providing
evidence for neutrino oscillations and thus for physics beyond the
Standard Model. This has led to new insights into the possibilities
for new physics, but while some old questions have been answered, many
new questions have emerged \cite{aps}.

An important question that still needs to be resolved is the
determination of the neutrino mass hierarchy. A very large effort has
been dedicated to planning new experiments for measuring all neutrino
oscillation parameters \cite{osc}. The mass hierarchy can be
determined using matter effects on oscillations inside the Earth. This
however requires a long baseline, a very large detector and an intense
beam. In addition, parameter degeneracies have to be resolved using a
combination of experiments.

Cosmic ray interactions in the atmosphere give a natural beam of
neutrinos.  These atmospheric neutrinos in the GeV range have been
used by the Super-Kamiokande detector to provide evidence for neutrino
oscillations. The large size of neutrino telescopes such as AMANDA,
IceCube and KM3NeT makes possible the detection of a large number of
atmospheric neutrino events with a higher energy threshold, $\sim
100$~GeV, even though the neutrino flux decreases rapidly with energy
($\sim E_\nu^{-3}$).  Built to detect neutrinos from astrophysical
sources, dark matter annihilation, etc. \cite{telescopes}, these
ice/water Cherenkov detectors typically have high detection threshold
energy to avoid the large background from atmospheric neutrinos.
Since neutrino oscillation effects are quite small at high energies,
high energy atmospheric neutrinos provide little information about
important issues such as neutrino mass hierarchy, mixing angles, etc.

Recently, a low energy extension of the IceCube detector has been
planned \cite{deepcore}.  Its goal is to significantly improve the
atmospheric muon rejection and to extend the IceCube neutrino
detection capabilities in the low energy domain, possibly to muon
energies as low as 5 GeV, depending on the density of the photo
tubes. The proposed instrumented volume is 5,000-10,000 kton by
conservative estimates. Such a low threshold array buried deep inside
IceCube will open up a new energy window on the universe.  It will
search for neutrinos from sources in the Southern hemisphere, in
particular from the galactic center region, as well as for neutrinos
from WIMP annihilation, as originally motivated.  Here we provide an
additional and independent motivation for building such an array,
namely to explore neutrino oscillation physics.

In this article we show that the deep core extension of IceCube
provides a great opportunity for detailed oscillations studies of
atmospheric neutrinos and makes possible the determination of the
neutrino mass hierarchy. This is extremely important given that long
baseline experiments with comparable sensitivity might take a very
long time to build and collect data, while the IceCube deep core will
be built in the near future and will accumulate high statistics
relatively fast. Our study indicates that for a total mass of the
instrumented volume times exposure of 100 Mt yr (roughly
equivalent to 20 years of running a 5,000 kton detector, or 10 years
of running a 10,000 kton detector in an optimistic scenario), 
neutrino mass hierarchy can be determined at least with $90\%$
confidence level assuming the current best-fit values of the
oscillation parameters, and for values of $\theta_{13}$ close to the
present bound.

We start in section \ref{sec:oscrev} with a brief review of what we
know about neutrino oscillation parameters and what we expect to learn
in the near future. We also discuss matter effects inside the Earth,
which play an important role in the analysis. In section
\ref{sec:analy} we describe our main analysis and in section
\ref{sec:backs} we discuss the backgrounds and systematic
uncertainties that affect this analysis. We present the main results
in section \ref{sec:results} and discuss them in section
\ref{sec:disc}.

\section{Neutrino Oscillations}
\label{sec:oscrev}

Neutrino data from solar, atmospheric, reactor and accelerator
experiments is well understood in terms of three-flavour neutrino
oscillations. Two $\Delta m^2$ values and two (large) mixing angles
are well determined, while the third mixing angle is limited to be
very small. The CP-violating phase ($\delta$) is completely
unconstrained.  In addition, the sign of $\Delta m^2_{31}$ is also
unknown and will be the focus of our study. The two possibilities,
$\Delta m^2_{31}>0$ or $\Delta m^2_{31}<0$ correspond to two types of
neutrino mass ordering, normal hierarchy and inverted hierarchy.

The best fit {oscillation} parameter values obtained from present data
are \cite{osc,sv}: 
\ba 
|\Delta m^2_{31}|&=&2.5\times 10^{-3} {\rm eV}^2\\ \Delta
m^2_{21}&=& 8 \times 10^{-5} {\rm eV}^2\\ \sin^22\theta_{23}&=&
1\\ \tan^22\theta_{12}&=& 0.45 \ea with 99\% CL allowed regions given
by: \ba |\Delta m^2_{31}|&\in&(2.1-3.1) \times 10^{-3} {\rm
  eV}^2\\ \Delta m^2_{21}&\in&(7.2-8.9) \times 10^{-5} {\rm
  eV}^2\\ \theta_{23} &\in & (36^\circ -54^\circ)\\ \theta_{12} &\in &
(30^\circ - 38^\circ)
\label{eq:paramunc}
\ea
and $\sin^22\theta_{13}\le0.15$ for $\Delta m^2_{31}=2.5\times 10^{-3}
{\rm eV}^2$. Notice that an extra unknown in the neutrino oscillation
scenario is the octant in which $\theta_{23}$ lies, if $\sin^2 2
\theta_{23}\ne 1$. This has been dubbed in the literature as the
$\theta_{23}$ octant ambiguity.

In the near future, long baseline experiments like MINOS and OPERA
will improve the current precision on $\Delta m^2_{31}$ and possibly
discover a non-zero value of $\theta_{13}$, if this is close to the
present upper limit. In a few years, reactor experiments like
DoubleChooz and DayaBay will provide improved sensitivity to
$\theta_{13}$. This information can be used as input in our analysis,
reducing some of the parameter uncertainties.

In the past, atmospheric neutrinos in the Super-Kamiokande detector
have provided evidence for neutrino oscillations and the first
measurements of $|\Delta m^2_{31}|$ and $\sin^22\theta_{23}$. It is
also known that, covering a large range of energies and pathlengths
and using matter effects inside the Earth, they can be in principle
sensitive to sub-dominant neutrino oscillation effects like
$\theta_{13}$ and the mass hierarchy
\cite{pepe1,fatis,pepe2,pepe3,rest,huber,inoteam,recent}.

Matter effects~\cite{matter,matterosc} in long baseline and
atmospheric neutrino oscillation experiments depend on the size of the
mixing angle $\theta_{13}$ which governs the transitions $\nu_e
\leftrightarrow \nu_{\mu,\tau}$ driven by the atmospheric mass squared
difference $\Delta_{31}=\Delta m^2_{31}/2E$. The effective
$\theta_{13}$ mixing angle in matter in a two-flavour framework is
given by:
\begin{equation}
\sin^2 2 \theta^{\textrm{m}}_{13}= \frac{\sin^2 2 \theta_{13}}{\sin^2
  2 \theta_{13} + \left(\cos 2 \theta_{13} \mp \frac{\sqrt{2} G_{F}
    N_{e}}{\Delta_{31}}\right)^2}~,
\label{eq:mixmatter}
\end{equation}
where the minus (plus) sign refers to neutrinos (antineutrinos), $N_e$
is the electron number density in the medium, $\sqrt{2} G_{F}
N_e$~(eV)$=7.6 \times 10^{-14} Y_e \rho$~(g/cm$^3$) and $Y_e$, $\rho$
the electron fraction and the density of the medium, the Earth
interior in our case. The electron fraction $Y_e$ is $0.466$ ($0.494$)
in the core (mantle) and we follow the PREM~\cite{prem} model for the
Earth's density profile.  Equation~(\ref{eq:mixmatter}) implies that,
in the presence of matter effects, the neutrino (antineutrino)
oscillation probability gets enhanced if the hierarchy is normal
(inverted). Making use of the different matter effects for neutrinos
and antineutrinos seems therefore the ideal way to distinguish among
the two possibilities: normal versus inverted mass hierarchy. Matter
effects are expected to be important when the resonance condition:
\begin{equation}
\Delta m^2_{31} \cos\left(2 \theta_{13}\right) = 2 \sqrt{2} G_{F}
N_{e} E~,
\label{eq:res}
\end{equation}
is satisfied. The precise location of the resonance will depend on
both the neutrino path and the neutrino energy. For $\Delta m^2_{31}
\sim 2.5 \times 10^{-3}$~eV$^2$ and distances of several thousand
kilometers the resonance effect is expected to take place for neutrino
energies $\mathcal{O}(10)$~GeV. The pathlength traveled by atmospheric
neutrinos is:
\begin{equation}
L(c_\nu) = R_\otimes ( \sqrt{(1+l/R_\otimes)^2-s_\nu^2}-c_\nu),
\end{equation} 
where $R_\otimes=6371$~km is the radius of the Earth and $l\sim 15$ km
is the typical height at which neutrinos get produced in the
atmosphere.  The cosine and sine of the nadir angle of the incident
neutrino are denoted by $c_\nu$ and $s_\nu$, respectively. Since
\emph{upward going} ($c_\nu \to -1$) atmospheric neutrinos traverse
the dense core of the Earth, they provide an excellent tool to tackle
the neutrino mass ordering.

Indeed the idea of exploiting matter effects in atmospheric neutrino
oscillations to distinguish the type of hierarchy has been extensively
explored in the literature~\cite{pepe1,fatis,pepe2,pepe3,rest}. In
general, the former studies exploit muon calorimeter detectors, such
as MONOLITH~\cite{monolith}, MINOS~\cite{MINOS} or INO~\cite{INO} in
which the muon charge can be determined. The measurement of the number
of positive and negative muons in the $1-10$~GeV energy region allows
then for a direct extraction of the neutrino mass hierarchy, simply by
looking in which channel (neutrino or antineutrino) the signal, via
matter effects, is enhanced. However, it has been pointed out, and
carefully explored, that the detection of atmospheric neutrinos which
have crossed the Earth by future planned megaton water Cherenkov
detectors could also determine the neutrino mass hierarchy, provided
the mixing parameter $\sin^2 2 \theta_{13}$ is not very
small~\cite{huber,inoteam,recent}, even when these detectors do not
allow for a charge discrimination of the leptons. The detection of
\emph{low energy} neutrinos in a higher density photo-multiplier array
within the IceCube instrumented detector volume opens up the
possibility of not only exploring the atmospheric oscillation
pattern~\cite{AS01}, but also exploring how well the neutrino mass
hierarchy could be measured in the \emph{largest} water/ice Cherenkov
detector available in the near future. In the next section we present
the details of the analysis proposed here.  For our numerical
analysis, unless otherwise stated, we will use the best fit values
quoted earlier in this section for the oscillation parameters.

\section{Analysis}
\label{sec:analy}

The IceCube detector has the ability to measure separately the muon
tracks and electron/tau generated cascades, thus providing good
flavour identification in some energy ranges. This would be extremely
useful for an oscillation analysis, especially when searching for
sub-dominant effects. In particular, the atmospheric neutrino
sensitivity to the mass hierarchy comes from the matter effects on
$\nu_\mu\to\nu_e$ ($\nu_e \to \nu_\mu$) oscillations and previous
studies dealing with water Cherenkov detectors have used the electron
signal to extract this information. In the low energy range relevant
for neutrino oscillations however, it is extremely hard in IceCube
(and even in the deep core array) to obtain information about neutrino
direction and energy for electron cascades.

We thus focus here on the $\mu$-like {\it contained} events produced
by the interactions of atmospheric upward going neutrinos in deep ice.
Formally, the expected number of muon neutrino-induced contained
events in the $i$- and $j$-th energy and cosine of the nadir angle
($c_\nu$) bins read:
\begin{widetext}
\begin{equation}
N_{i,j,\mu}= \frac{2 \pi N_{\rm T} \, t}{V_{\rm det}} \,
\int_{E_i}^{E_i +\Delta_i}dE_\nu \int_{c_{\nu, j}}^{c_{\nu,j}
  +\Delta_j}dc_\nu V_\mu\times
\left(\frac{d\phi_{\nu_\mu(\nu_e)}}{dE_\nu d\Omega}\, \sigma^{\rm
  CC}_{\nu_\mu(\nu_e)} P_{\nu_\mu(\nu_e) \to \nu_\mu}
+\frac{d\phi_{\bar{\nu}_\mu(\bar{\nu}_e)}}{dE_\nu d\Omega}\,
\sigma^{\rm CC}_{\bar{\nu}_\mu(\bar{\nu}_e)}
P_{\bar{\nu}_\mu(\bar{\nu}_e)\to \bar{\nu}_{\mu}}\right)~,
\label{eq:events}
\end{equation}
\end{widetext}
where $\Delta_i$ and $\Delta_j$ are respectively the energy and
$c_\nu$ bin widths, $N_{\rm T}$ is the number of available targets,
$V_{\rm det}$ is the total volume of the detector, $t$ is the exposure
time, $d\phi_\nu$'s are the atmospheric (anti)neutrino differential
spectra, $\sigma^{\rm CC}$ is the CC (anti)neutrino cross section and
$V_{\mu}$ is the effective detector volume.  For a detector with
cylindrical shape of radius $r$ and height $h$, $V_\mu$ is given
by~\cite{AS01}
\begin{widetext}
\beq
V_\mu(E_\mu,\theta)= 2 h r^2
\arcsin\left(\sqrt{1-\frac{R^2_\mu(E_\mu)}{4 \, r^2}\sin^2 \theta}
\right) \left(1-\frac{R_\mu(E_\mu)}{h}|\cos\theta|\right)~, 
\label{eq:range}
\enq
\end{widetext}
where $R_\mu(E_\mu)$ is the energy-dependent muon range in ice.  For
$\sigma^{CC}$ we use the charged current (anti)neutrino interaction
cross-sections in \cite{GQRS95}. It is useful to note that in the
relevant energy range the anti-neutrino cross-section is smaller than
the neutrino cross-section by about a factor of two. This difference
is what allows for the (statistical) discrimination between neutrinos
and anti-neutrinos and thus between normal and inverted hierarchy in
this experiment.

Equation~(\ref{eq:events}) contains the atmospheric electron and muon
(anti) neutrino fluxes, $\frac{d\phi_{\nu_\alpha}}{dE_\nu
  d\Omega}$. For the results presented in this study we use the
results from Refs.~\cite{fluxesus}. The atmospheric neutrino fluxes
from Refs.~\cite{hondaetal} have also been used, and, overall, we
obtain a similar difference between the number of muon neutrino
induced events for the normal and inverted hierarchies both in energy
and $c_\nu$. The absolute electron and muon atmospheric (anti)neutrino
fluxes are found to have errors of $10\%-15\%$ in the energy region of
interest here~\cite{uncertainties}. Those errors are mostly induced by
our ignorance in modeling hadron production, although the situation is
expected to improve with HARP and MIPP data. The uncertainties quoted
above are reduced for the neutrino-antineutrino flavor ratio
case~\footnote{In general, the different available computations of the
  atmospheric neutrino fluxes~\cite{fluxesus,hondaetal} predict almost
  the same neutrino-antineutrino ratios.}, i.e. for
$\nu_\mu/\bar{\nu}_\mu$ and $\nu_e/\bar{\nu}_e$, where the uncertainty
is $\sim 7\%$ in the energy range we explore in the current
study. Even smaller uncertainties are expected when the
muon-to-electron flavor ratio $(\nu_\mu +
\bar{\nu}_\mu)/(\nu_e+\bar{\nu}_e)$ is considered. We will comment on
the impact of the atmospheric neutrino flux uncertainties on our
results below, when including systematic uncertainties to our
numerical analysis.

\begin{figure*}[t]
\includegraphics[width=2.3in]{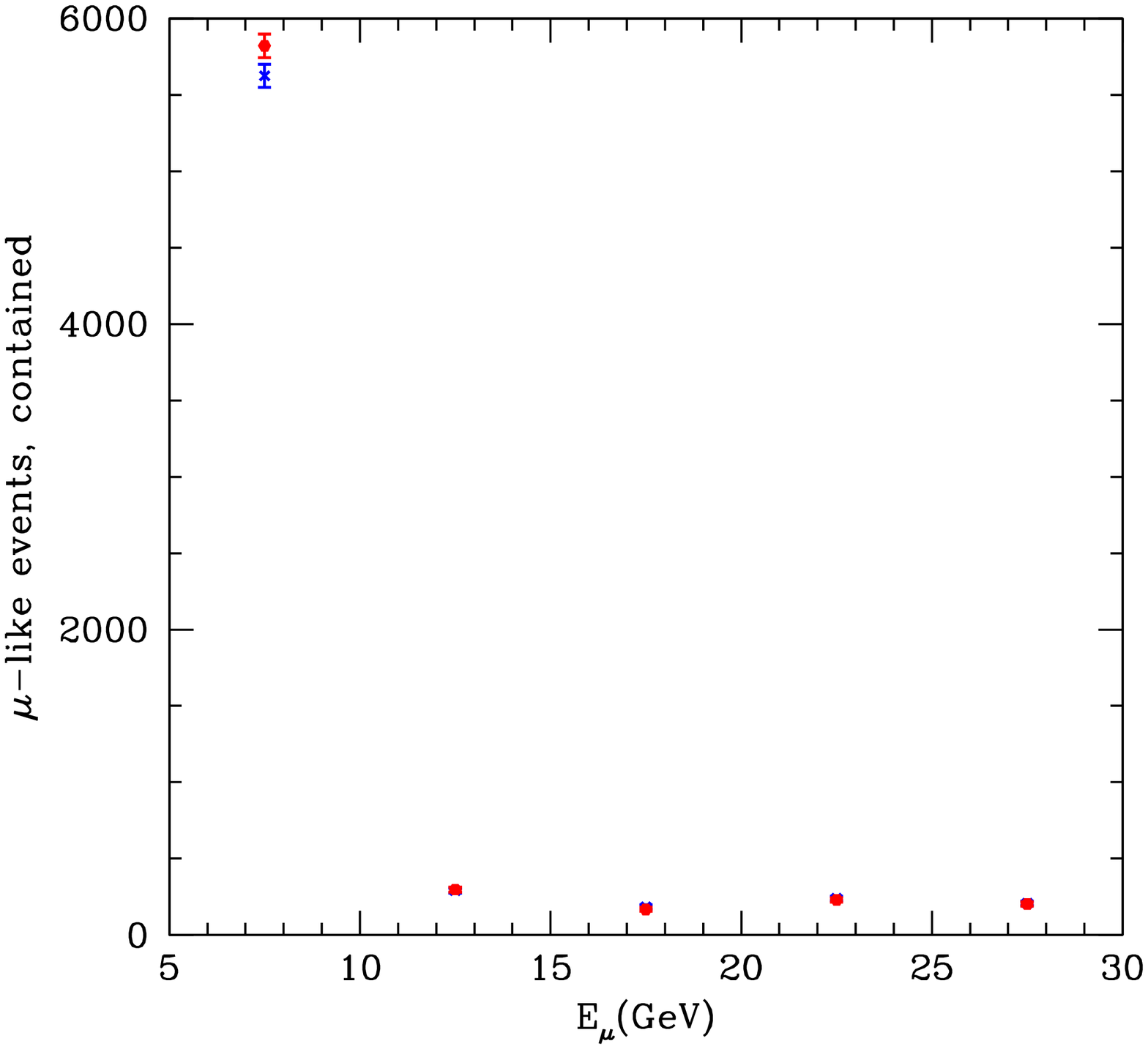}
\includegraphics[width=2.3in]{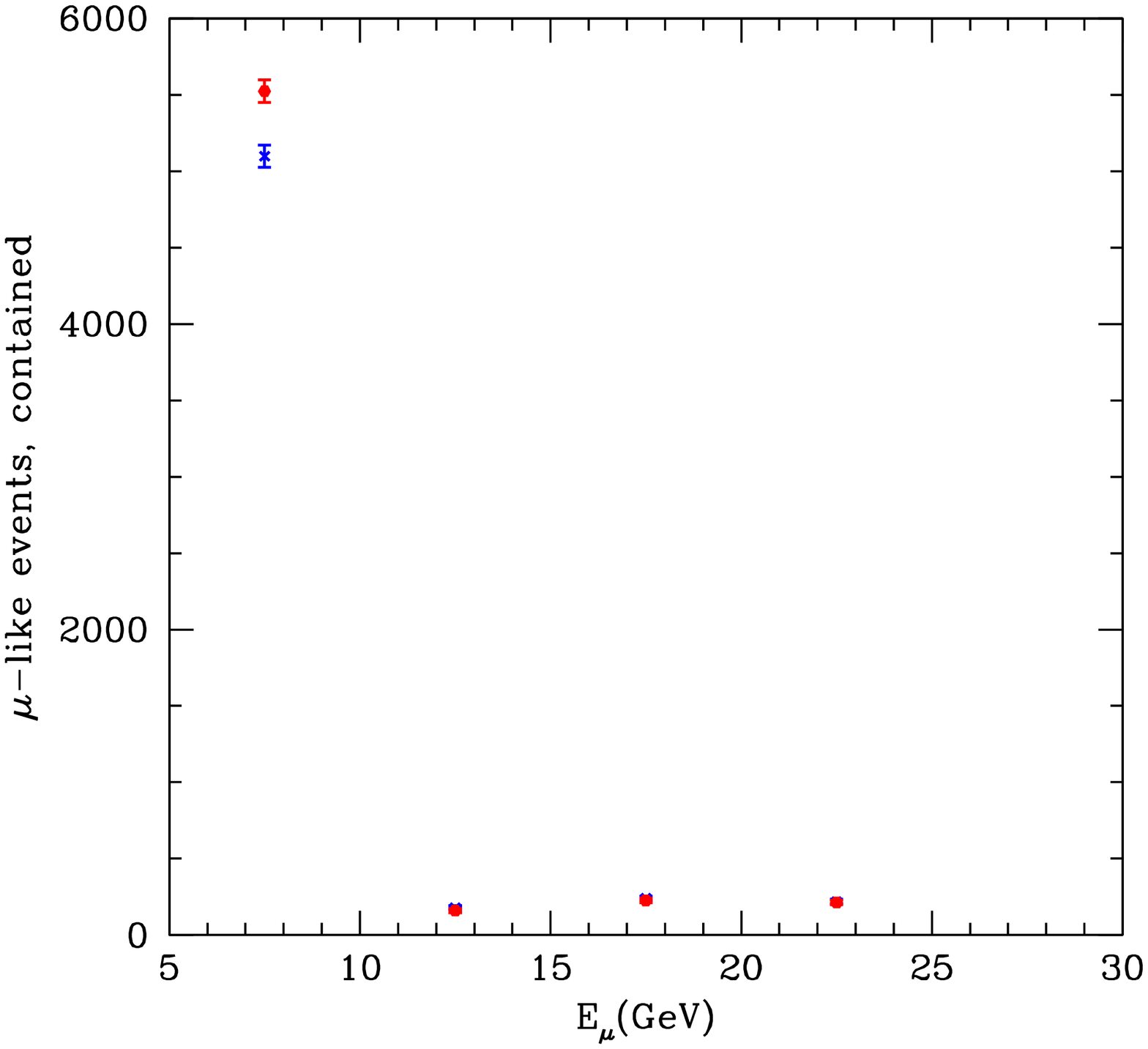}
\includegraphics[width=2.3in]{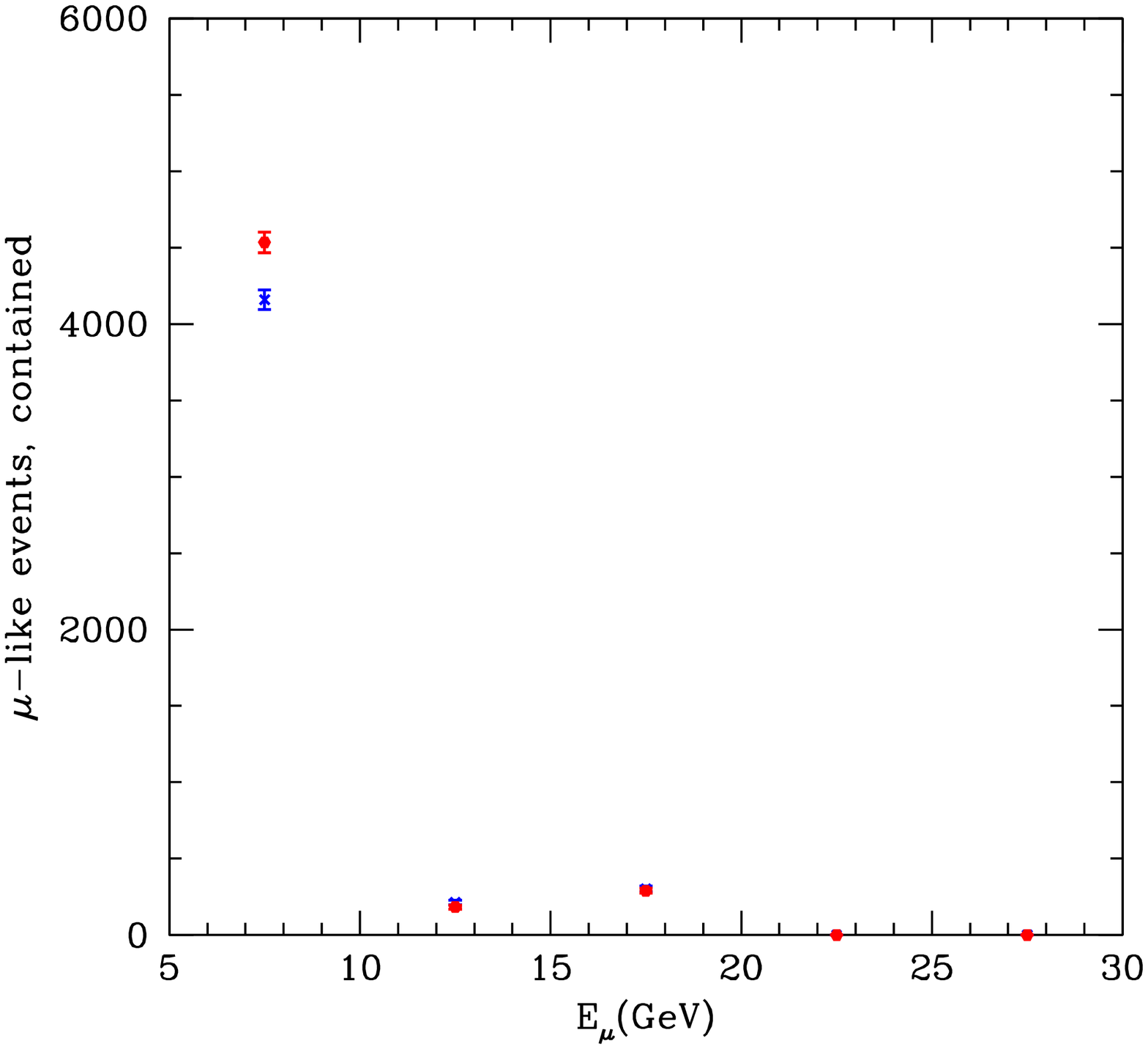}
\caption{From left to right, number of contained $\mu$-like events in
  a $5000$~kton detector after 10 years exposure, within the
  $(-1,-0.9)$, $(-0.9,-0.8)$ and $(-0.8,-0.7)$ $c_\nu$ bin, assuming
  $\sin^2 2\theta_{13}=0.1$, $\delta=0$ and an energy bin size of
  $5$~GeV, with a muon energy threshold for detection of $5$~GeV. We
  show the $1\sigma$ statistical errors. The blue crosses (red
  circles) denote normal (inverted) hierarchy.}
\label{fig:fig1}
\end{figure*}

The energy of secondary muons from CC interaction in the 10-100 GeV
neutrino energy range of interest here is $\langle E_\mu \rangle =
0.52 \, E_\nu$ and $0.66 \, E_\nu$, respectively for neutrinos and
antineutrinos~\cite{GQRS95}. We illustrate in Fig.~\ref{fig:fig1} the
expected $\mu$-like \emph{contained} events in 5 GeV \emph{muon
  energy} bins for a combined detector mass times exposure of 50 Mt
yr.  From left to right, the panels depict the contained $\mu$-like
events within the $(-1,-0.9)$, $(-0.9,-0.8)$ and $(-0.8,-0.7)$ $c_\nu$
bins. We assume $\sin^2 2\theta_{13}=0.1$ and $\delta=0$ along with
other best-fit parameters in Eq. (\ref{eq:paramunc}).  Although we
used a detector geometry of $1$~km height and $\sim 40$~m radius, all
events are contained in these $c_\nu$ bins except for the highest
energy bin: (25,30)~GeV. As we will shortly see, the oscillation
signals solely affect the low energy events. Thus our results are
valid for a variety of detector geometries, only affected by the total
instrumented volume times the exposure.

For the first $c_\nu$ bin used in Fig.~\ref{fig:fig1} the
(anti)neutrinos are almost vertically upward going and in their way
towards the detector they have crossed the high density Earth
core. For the other bins (anti)neutrinos still traverse a significant
amount of matter through the Earth. Note that a finer angular bin than
$\Delta c_\nu \sim 0.1$ is not possible because the reconstruction of
primary neutrino direction is expected to be poor in this energy range
due to the intrinsic spread in charged lepton-neutrino scattering
angle.  The resonance is expected to be located at low energies, see
Eq.~(\ref{eq:res}), and the maximum difference between normal and
inverted hierarchies is observed in the $(5,10)$~GeV energy bin. In
higher energy bins, the effect is totally negligible.

\begin{figure*}[t]
\includegraphics[width=3in]{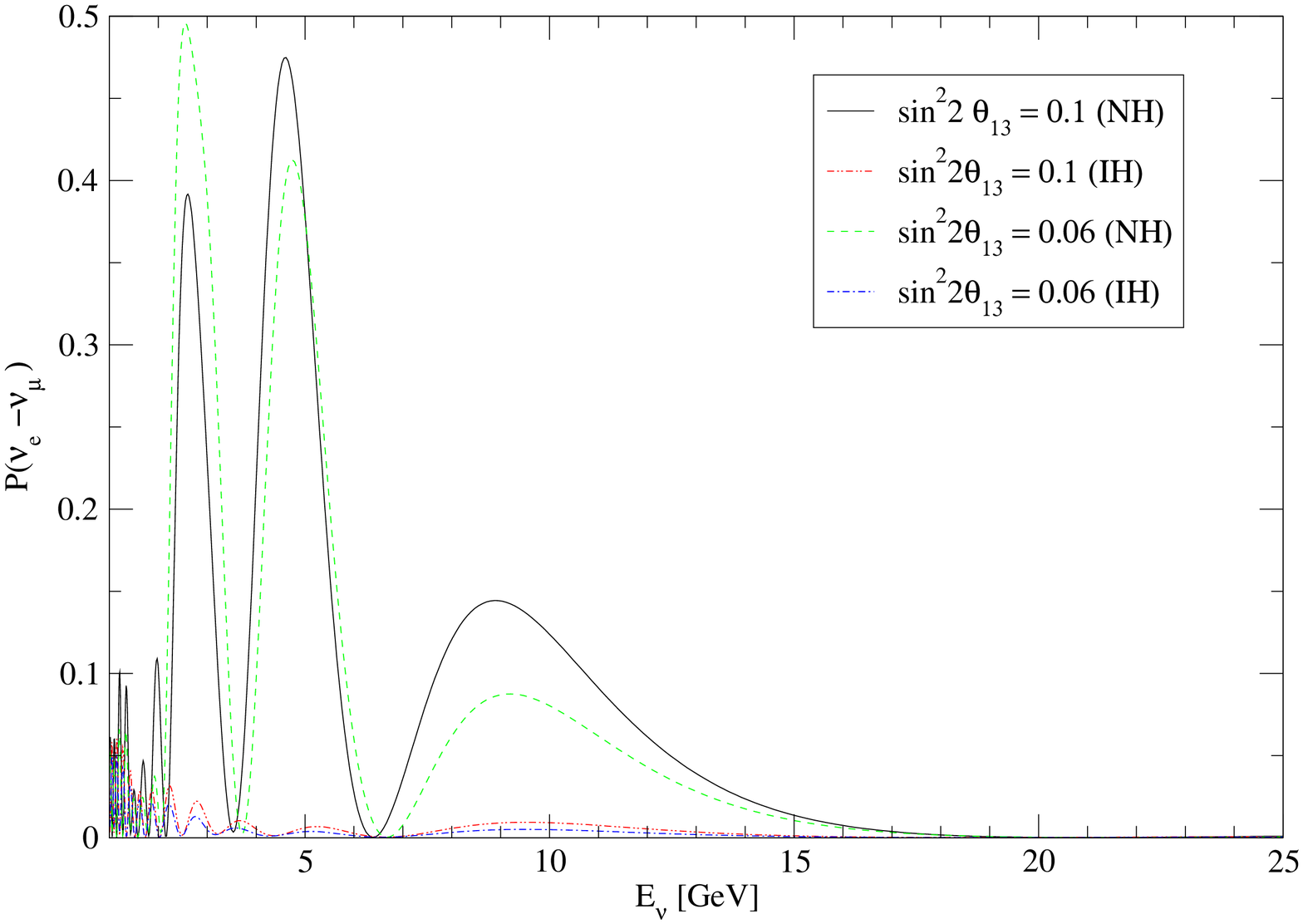} \hskip 0.1in
\includegraphics[width=3in]{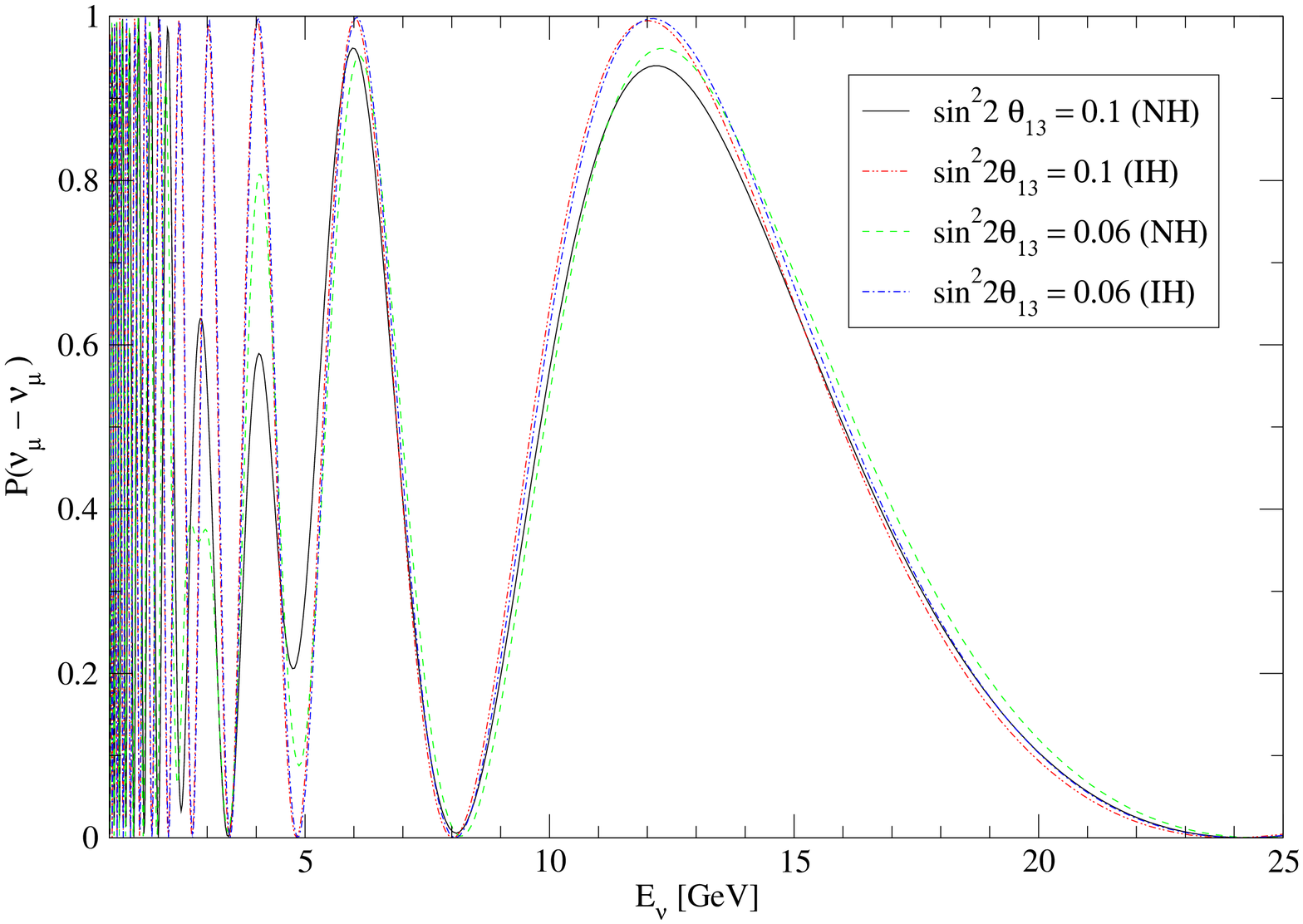}
\caption{left (right panel): Oscillation probabilities for
$\nu_e\to\nu_\mu$, $\nu_\mu\to\nu_\mu$ transitions for $c_\nu=-1$.}
\label{fig:probs}
\end{figure*}

Figure~\ref{fig:fig1} helps understanding the energy and angular bins
that should be considered for the numerical analysis developed in the
current study. We will exploit exclusively the first three muon energy
bins (i.e. $E_\mu$ within the $(5,10)$~GeV, $(10,15)$~GeV and
$(15,20)$~GeV energy ranges), and three angular bins, i.e., $c_\nu$
within the $(-1,-0.9)$, $(-0.9,-0.8)$ and $(-0.8,-0.7)$ ranges).  It
is precisely in those bins where the sensitivity to the neutrino mass
ordering is significant, and all the muon events are fully contained.
Since the atmospheric neutrino spectra are very steep ($\sim
E_\nu^{-3}$ in the relevant energy range) and, as we will discuss
next, this energy range also corresponds to a neutrino oscillation
maximum, most events are concentrated in the lowest energy bin. It is
thus important to keep the muon energy threshold for detection as low
as possible (between $5$ and $7$~GeV) to collect large statistics as
well as for extracting the neutrino mass ordering.  Note that, as
discussed above, 5-7 GeV muon energy threshold corresponds to
approximately 10-15 GeV initial neutrino energy. With a proposed 7 m
spacing between phototubes on the deep core strings, such a low energy
threshold is achievable. We next discuss the oscillation effects in
the event rates in Fig.~\ref{fig:fig1}.  Notice that for the three
angular bins illustrated here, the number of $\mu$-like events in the
first energy bin is larger for the inverted hierarchy case than for
the normal hierarchy case. The reason for that is due to the presence
of matter effects at these zenith angles for $\nu_\mu$'s, which are
the ones dominating the statistics. This can be well seen in
Fig.~\ref{fig:probs}, which shows $P_{\nu_e\to\nu_\mu}(\equiv
P_{e\mu})$ and $P_{\nu_\mu\to\nu_\mu} (\equiv P_{\mu\mu})$ for both
normal and inverted hierarchy, for $\sin^22\theta_{13}=0.1$ and
$\sin^22\theta_{13}=0.06$.

If the hierarchy is normal (inverted), the $P_{\mu \mu}$ survival
probabilities, in the $\sim$10-20 GeV energy range for neutrinos
(which are responsible to produce 5-10 GeV muons), are suppressed
(enhanced), due to matter effects, and therefore a smaller (larger)
number of $\nu_\mu$ CC interactions are expected in the detector. If
the matter density is constant, and the contribution from the solar
terms is negligible, the $P_{\mu \mu}$ survival probability is given
by (see Refs.~\cite{choubey,recent})
\ba
P_{\mu \mu}&=& 1-\cos^2 \theta^{\textrm{m}}_{13} \sin^2 2
\theta_{23}\\ \nonumber &&\times \sin^2\left[1.27\left(\frac{\Delta
    m^2_{31} +A + (\Delta
    m^2_{31})^{\textrm{m}}}{2}\right)\frac{L}{E}\right]\\ \nonumber
&&-\sin^2 \theta^{\textrm{m}}_{13} \sin^2 2 \theta_{23}\\ \nonumber
&&\times \sin^2\left[1.27\left(\frac{\Delta m^2_{31} +A - (\Delta
    m^2_{31})^{\textrm{m}}}{2}\right)\frac{L}{E}\right]\\ \nonumber
&&- \sin^4\theta_{23}\sin^2 2 \theta^{\textrm{m}}_{13}
\sin^2\left[1.27 (\Delta m^2_{31})^{\textrm{m}}\frac{L}{E}\right]~,
\label{eq:mixmatter2}
\ea
where $A=2 \sqrt{2} G_{F} N_{e} E$, $\theta^{\textrm{m}}_{13}$ is given by 
Eq.~(\ref{eq:mixmatter}) and 
\begin{equation}
(\Delta m^2_{31})^{\textrm{m}}= \Delta m^2_{31} \sqrt{\sin^2 2
    \theta_{13} + \left( \cos 2 \theta_{13} \mp \frac{2 \sqrt{2} G_{F}
      N_{e} E}{\Delta m^2_{31}} \right)^2}~,
\label{eq:masseff}
\end{equation}
where the minus (plus) sign applies to neutrino (antineutrino) flavor
transitions. Due to the presence of resonant matter effects in the
first angular and energy bins, $P_{\mu \mu}$ can be very different for
normal and inverted hierarchies (mostly due to changes in the first
term in Eq.~(\ref{eq:mixmatter2})).

As seen in Fig.~\ref{fig:probs}, the $\nu_e\to \nu_\mu$ probability is
quite small in the relevant energy range (negligible for inverted
hierarchy). Its contribution to the final muon-like event rate is made
even smaller by the fact that the atmospheric $\nu_e$ flux at these
energies is much smaller than the $\nu_\mu$ flux. The experiment we
are considering is thus mostly exploiting matter effects in the {\em
  disappearance} $\nu_\mu\to\nu_\mu$ channel and therefore is in many
ways complementary to the appearance experiments. While the matter
effects are a small correction in the $\nu_\mu$ survival probability,
they are sufficient to provide a difference between the different mass
orderings because of the very large number of events.

Note that in Fig.~\ref{fig:fig1} the difference between event rates
for the two hierarchies increases (although the overall rates
decreases) for $c_\nu$ bins $(-0.9,-0.8)$ and $(-0.8,-0.7)$ compared
to the $(-1,-0.9)$ bin. This is because the resonant matter density
for neutrino energies in the first energy bin $<E_\nu>=15$~ GeV is
$\sim$5 g/cm$^3$ which is lower than the densities that the neutrino
crosses if $c_\nu$ is in the $(-1, -0.9)$ region, but gets closer to
the ones in the shallower $c_\nu$ region.

\section{Backgrounds and systematic uncertainties}
\label{sec:backs}

The main backgrounds to the signal we are exploiting in the current
study are atmospheric downward going muons from the interactions of
cosmic rays in the atmosphere and tau (anti)neutrinos from
$\nu_{\mu,e}(\bar{\nu}_{\mu,e}) \to \nu_\tau(\bar{\nu}_\tau)$
transitions.  The cosmic muon background can be eliminated by angular
cuts and in the Ice Cube deep core is significantly reduced compared
to the IceCube detector.

The tau neutrino background can be included in the analysis as an
additional source of $\mu$-like events. Tau (anti)neutrinos resulting
from atmospheric neutrino flavor transitions will produce a $\tau$
lepton by CC interactions in the detector effective volume.  The tau
leptons produced have an $\sim 18\%$ probability of decaying through
the $\tau^{-}\to \mu^{-}\bar{\nu}_\mu \nu_\tau$ channel.

\begin{figure} [ht]
\centerline{\epsfxsize=3in \epsfbox{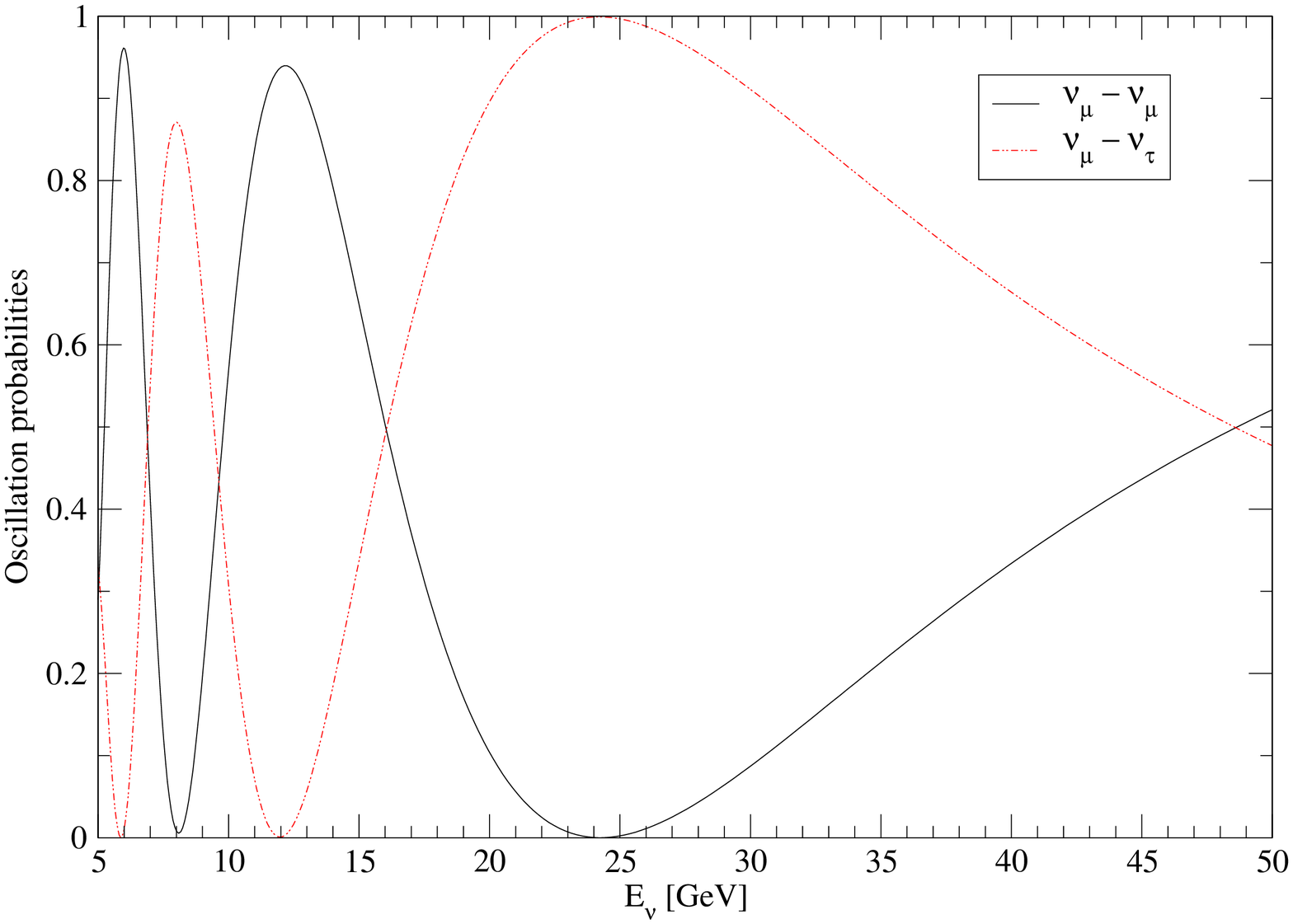}} 
\caption{$\nu_\mu$ survival probability and $\nu_\mu\to\nu_\tau$
  oscillation probability for $c_\nu=-1$, $\sin^22\theta_{13}=0.1$}
\label{fig:probtau}
\end{figure}

The secondary muons can mimic muons from $\nu_\mu$ CC interactions and
must be included in the oscillated signal. The energy of a $\nu_\tau$
needs to be about 2.5 times higher than a $\nu_\mu$ to produce, via
tau decay, a muon of the same energy. But the atmospheric neutrino
flux has a steeply falling spectrum, so one would expect this
tau-induced muon background not to be very large. It is however
significant ($\sim 10\%$) due to the fact that, as seen in
Figure~\ref{fig:probtau}, the first maximum in the
$\nu_\mu\to\nu_\tau$ oscillation probability (minimum in the
$\nu_\mu\to\nu_\mu$ survival probability) falls exactly in the energy
range of interest and for a large range the $\nu_\tau$ flux can be
significantly larger than the $\nu_\mu$ flux.  These events
significantly change the energy spectrum of the measured muon-like
events and contain information about the main oscillation parameters,
$\Delta_{31}$ and $\theta_{23}$.

The uncertainties in the atmospheric neutrino flux have been discussed
in the previous section and they affect the analysis. It is however
possible to use the data itself to improve some of the errors
introduced by these effects, by considering energy and angular bins
where oscillation effects are not important as a reference and thus
canceling out some of these uncertainties in the analysis (see also
\cite{atmflux}).

The uncertainties in other oscillation parameters also affect the
possibility of determining the neutrino mass hierarchy. We employ a
full three-flavour oscillation analysis. It is however easy to
understand that the solar $\Delta m^2$ and mixing angle have almost no
contribution. This is due to the high energy values we are
considering, for which even the largest distances traveled by
atmospheric neutrinos do not allow the $\Delta m^2_{21} L/E$ term to
become significant. The present uncertainties in $|\Delta m^2_{31}|$
and $\theta_{23}$ (see Eq. (\ref{eq:paramunc})) will be improved by
present accelerator experiments. In addition, the atmospheric neutrino
data in the IceCube deep core can be used to extract these parameters
independently from the sub-dominant effects, by comparing different
angular bins. The values of $\theta_{13}$ and $\delta$ are the most
uncertain and usually hardest to disentangle from the mass
hierarchy. A measurement of the $\theta_{13}$ mixing angle by the near
future reactor experiments would guarantee the possibility of
extracting the mass hierarchy from the atmospheric neutrino data in
IceCube. We consider in our analysis a range of values for
$\theta_{13}$ and $\delta$ in order to assess the parameter space for
which the neutrino mass hierarchy can be determined.  It is important
to emphasize however that the sensitivity to the CP-violating phase
$\delta$ and the value of $\theta_{23}$ in this experiment is quite
small due to the fact that the signal is dominated by the $\nu_\mu$
{\em disappearance} channel.

\section{Results}
\label{sec:results}

In this section we present our results in terms of a minimum $\chi^2$
analysis in the $(\sin^2 2\theta_{13}; \delta)$ parameter space.  For
a particular hierarchy (h) and $(\sin^2 2\theta_{13}; \delta)$
parameters chosen by nature, we consider the number of $\mu$-like
events $N^{\rm ex}_{ij,h} (\sin^2 2\theta_{13}; \delta)$ measured by
an experiment in the $i$- and $j$-th muon energy and $c_\nu$ bins (see
Eq.~(\ref{eq:events})). These events include the $\nu_\mu$ and
$\bar{\nu}_\mu$ signal, as well as the background secondary muons from
$\nu_\tau$ and $\bar{\nu}_\tau$'s.

The $\chi^2$ statistics, for a
``theoretical'' model of hierarchy ($h'$) and parameters $(\sin^2
2\theta'_{13}; \delta')$ is then defined as
\ba
\chi^2_{h'} (\sin^2 2\theta'_{13}; \delta') =
\sum_{i=1,3}\sum_{j=1,3} \,\,\,\, \nonumber \\ \left[ 
\frac{N^{\rm ex}_{ij,h} (\sin^2 2\theta_{13}; \delta) -
N^{\rm th}_{ij,h'}(\sin^2 2\theta'_{13}; \delta')}
{\sigma^{\rm ex}_{ij,h} (\sin^2 2\theta_{13}; \delta)} \right]^2~.
\label{eq:chi2}
\ea
Here $N^{\rm th}_{ij,h'}(\sin^2 2\theta'_{13}; \delta')$ is the
expected event number from both signal $\nu_\mu$'s and background
$\nu_\tau$'s given a ``theoretical'' model. The variance $\sigma^{\rm
  ex}_{ij,h} (\sin^2 2\theta_{13}; \delta)$ is calculated from
experimental events with or without systematic uncertainties.  We
minimize $\chi^2$ in Eq.~(\ref{eq:chi2}) for $(\sin^2 2\theta'_{13};
\delta')$ parameters (i.e. 2 d.o.f). When nature's choice or ``true''
hierarchy is $h$ (normal for example), then the ``wrong'' theoretical
model of hierarchy $h' \ne h$ (inverted in this case) is rejected if
\beq {\rm min} 
\left( \chi^2_{h' \ne h} \right) -
{\rm min} \left( \chi^2_{h'=h} \right) \ge \alpha 
\enq 
in the $(\sin^2 2\theta_{13}; \delta)$ parameter space.  The $68\%$,
$90\%$, $95\%$ and $99\%$ confidence levels (CL) are defined for
$\alpha = $~2.3, 4.61, 5.99 and 9.21 respectively, for 2 d.o.f
statistics~\cite{pdgstat}.  Note that our choice of 2 d.o.f statistics
is rather conservative as explained in the Appendix of
Ref.~\cite{stat}.  Our $90\%$ CL translate into the $97\%$ CL for 1
d.o.f statistics.

\begin{figure*} [ht]
\includegraphics[height=3in]{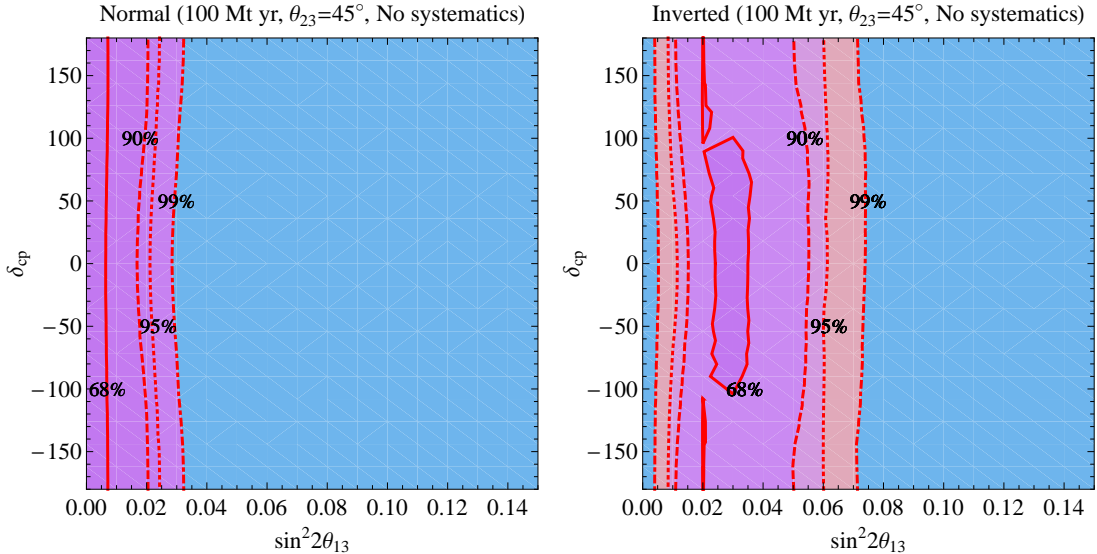}                          
\caption{ Rejection regions of the ``wrong'' hierarchy model in the
  $(\sin^2\theta_{13},\delta_{cp})$ plane when the ``true'' hierarchy
  is normal (left panel) or inverted (right panel) as indicated in the
  heading of each plot.  Different lines correspond to rejection
  regions of the ``wrong'' hierarchy at $68\%$, $90\%$, $95\%$ and
  $99\%$ CL (2 d.o.f.) using the muon-like contained events in a
  detector of mass times exposure of 100 Mt yr ($5,000$~kton detector after $20$ years of data taking, or $10,000$~kton
  detector after $10$ years of data taking in the optimistic
  scaenario). We used the best fit
  parameter values in Eqs.~(1)-(4). }
\label{fig:fig5}
\end{figure*}

Fig.~\ref{fig:fig5}, left (right) panel shows the $\chi^2$ results in
the ($\sin^2 2\theta_{13}, \delta$) plane for a measurement of the
hierarchy at the different confidence levels quoted above, exploiting
the muon-like contained events in a $5,000$~kton detector after $20$ 
years exposure (100 Mt yr), for normal (inverted) hierarchies. The
sensitivity to the mass hierarchy is better in case nature has chosen
the normal hierarchy than in the case for the inverted hierarchy. The
reason for that is that in the normal mass ordering scenario, matter
effects decrease (increase) the muon neutrino (antineutrino) survival
probability. In the inverted mass ordering scenario, matter effects
decrease (increase) the muon antineutrino (neutrino) survival
probability. Notice the observable we are exploiting here contains
both the atmospheric neutrino and antineutrino fluxes, see
Eq.~(\ref{eq:events}). Since in the energy range of interest the
neutrino fluxes are roughly twice the antineutrino ones, the overall
effect for the normal hierarchy is larger, i.e. no significant
cancellation among the effects for neutrinos and antineutrinos is
present. In the inverted hierarchy case, there exists a partial
cancellation and therefore the sensitivity results are not as
promising as the ones for the normal hierarchy.

\begin{figure*} [ht]
\includegraphics[height=3in]{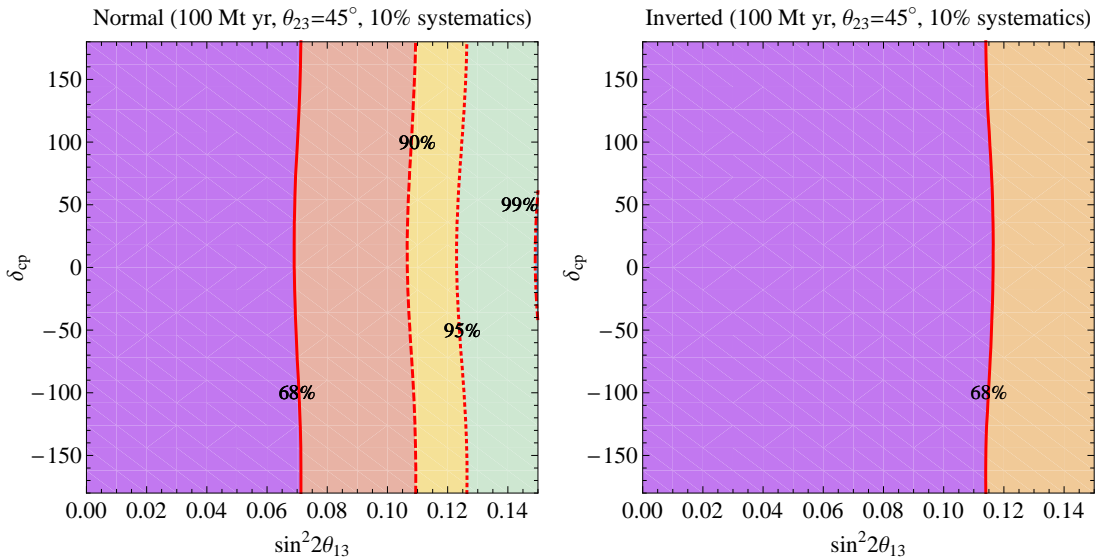}
\caption{Same as Fig.~\ref{fig:fig5} but including a $10\%$
    systematic error in our $\chi^2$ analysis.}
\label{fig:fig6}
\end{figure*}

The dependence of the $\chi^2$ results on the value of the CP-phase
$\delta$ is extremely mild, as expected, due to the fact that the muon
neutrino survival probability in matter $P_{\mu\mu}$,
Eq.~(\ref{eq:mixmatter2}), does not depend on $\delta$ in the limit of
negligible solar effects. Even when solar effects are not negligible,
$P_{\mu\mu}$ in matter depends exclusively on $\cos \delta$.
Consequently, one would expect the same results for $\delta$ and
$-\delta$, as can be clearly noticed from Figs.~\ref{fig:fig5}.  Also,
notice how the maximum (minimum) sensitivity is reached at
$\delta=0^\circ$ ($\delta=180^\circ$), due to the change of sign in
the $\cos \delta$ function.

Figs.~\ref{fig:fig6} show the equivalent to Figs.~\ref{fig:fig5} but
including in the $\chi^2$ analysis performed here a $10\%$ overall
systematic error, which includes uncertainties from atmospheric
neutrino fluxes and cross sections, among others. The impact of the
systematic errors on the sensitivity/exclusion curves is significant,
therefore, keeping systematic uncertainties below the $10\%$ level is
crucial for extracting the neutrino mass hierarchy.  While this value
might seem optimistic given our present knowledge of fluxes and
cross-sections, it is expected these uncertainties will be
significantly smaller on the time scale relevant for the IceCube deep
core experiment. This is likely to happen in two ways: a number of
precise cross-section measurements in the relevant energy range will
be performed in the near future, and, most importantly, the systematic
uncertainties can improve with data from the Deep Core itself. Our
analysis only makes use of the lowest energy range and almost straight
up-going neutrinos, which are most sensitive to the signal we are
interested in. Neutrinos from all other directions and at higher
energies are no longer sensitive to the sub-dominant neutrino
oscillation effects, but they contain important information about
fluxes, cross-sections and potentially main oscillation
parameters. They can thus be used as a ``reference'' in order to
minimize the systematic uncertainties for our analysis.

\begin{figure*} [ht]
\includegraphics[height=3in]{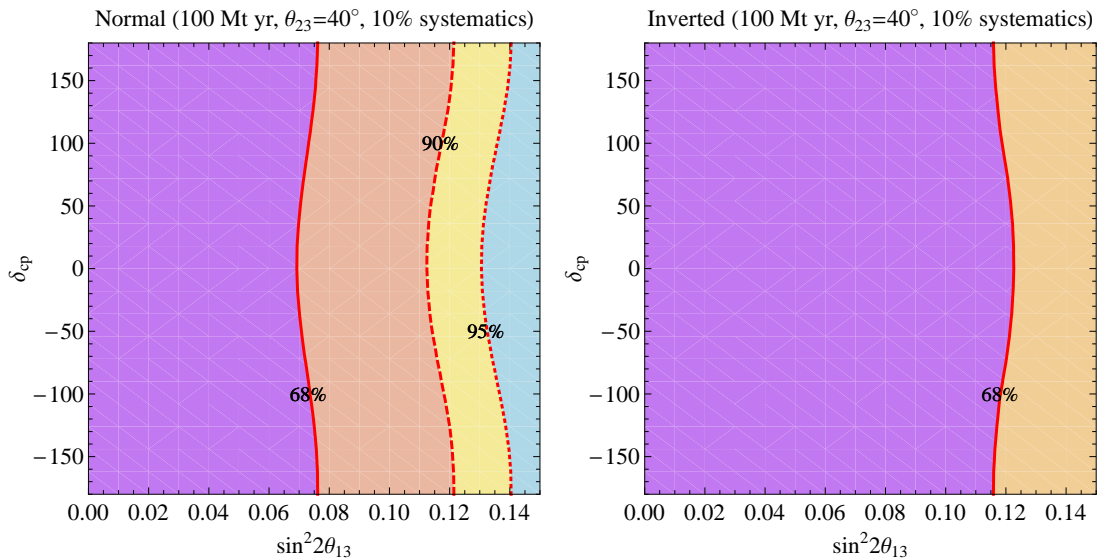}
\caption{ Left (right) panel: the different lines depict the $68\%$
  and $90\%$ CL (2 d.o.f.) hierarchy resolution using the muon-like
  contained events {in a detector with $100$~Mt~yr exposure,}
  for normal (inverted) hierarchies, and the
  atmospheric mixing angle chosen to be $\theta_{23}=40^\circ$. An
  overall $10\%$ systematic error has been included in the $\chi^2$
  analysis. }
\label{fig:fig7}
\end{figure*}

\begin{figure*} [ht]
\includegraphics[height=3in]{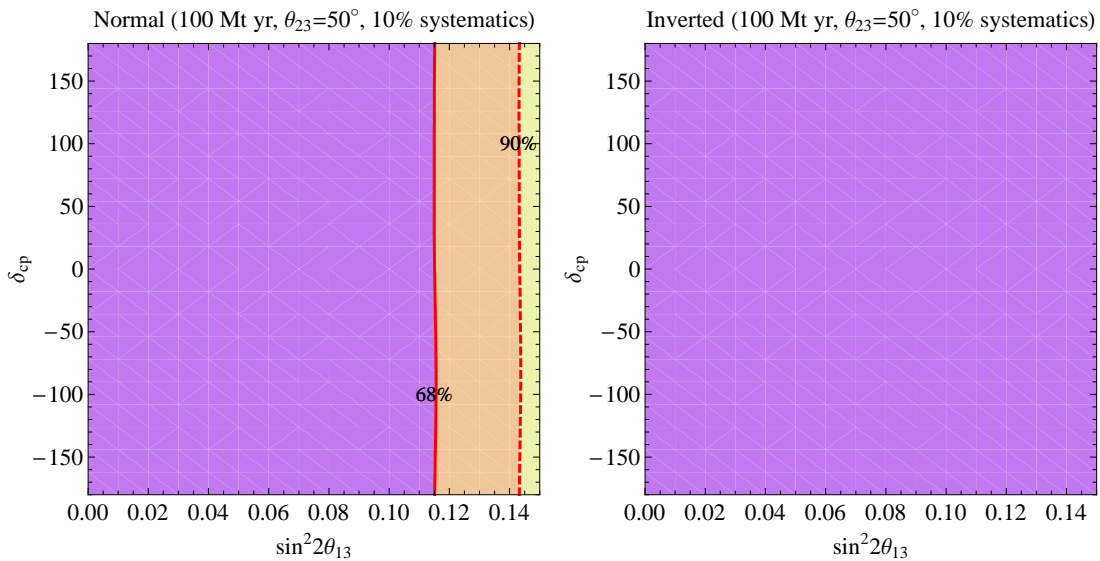}
\caption{ Left (right) panel: the different lines depict the $68\%$
  and $90\%$ CL (2 d.o.f.) hierarchy resolution using the muon-like
  contained events {in a detector with $100$~Mt~yr exposure,} for
  normal (inverted) hierarchies, and the
  atmospheric mixing angle chosen to be $\theta_{23}=50^\circ$. An
  overall $10\%$ systematic error has been included in the $\chi^2$
  analysis. }
\label{fig:fig8}
\end{figure*}

We also explore the impact of the $\theta_{23}$-octant
ambiguity. Following the current $95\%$ CL allowed limits for the
mixing parameter $\sin^2 2 \theta_{23}$, we illustrate here the
results from our $\chi^2$ analysis for $\theta_{23}=40^\circ$ and
$\theta_{23}=50^\circ$.  Figs.~\ref{fig:fig7} show the $\chi^2$
analysis results, including a $10\%$ overall systematic error for
$\theta_{23}=40^\circ$. Figs.~\ref{fig:fig8} show the equivalent but
for $\theta_{23}=50^\circ$.

If $\theta_{23} \neq 45^\circ$, the results for the normal hierarchy
become slightly worse, since $\sin^2 2 \theta_{23}$ is smaller than
unity and therefore the role of matter effects in the $P_{\mu \mu}$
oscillation probability is smaller (see Eq.~(\ref{eq:mixmatter2})).

For the inverted hierarchy case, the best sensitivities are reached
when $\theta_{23}=40^\circ$. This can be understood in terms of the
muon (anti) neutrino disappearance probability $P_{\mu \mu}$ in the
presence of matter effects and non negligible solar effects. If one
performs an expansion up to second order in the small parameters
$\Delta m^2_{21}/\Delta m^2_{31}$ and $\sin \theta_{13}$, there are
two terms in the muon (anti) neutrino disappearance probability
equation which depend on $\cos \delta$. One term is proportional to
$\sin 2 \theta_{23} \cos \delta$, the other one is proportional to
$\cos 2 \theta_{23} \cos \delta$, and therefore it is only different
from zero for $\theta_{23} \neq 45^\circ$, changing its sign
accordingly to the octant in which $\theta_{23}$ lies. For the muon
antineutrino channel and inverted hierarchy, this term, i.e. the term
proportional to $\cos 2 \theta_{23} \cos \delta$ is the dominant one
among the two terms proportional to $\cos \delta$ in the muon (anti)
neutrino disappearance probability equation.  For
$\theta_{23}<45^\circ (>45^\circ)$, the term proportional to $\cos 2
\theta_{23} \cos \delta$ maximizes (minimizes) the impact of matter
effects in $P_{\mu \mu}$. The former term is also the one responsible
for the change of shape in the sensitivity curves for
$\theta_{23}=40^\circ$ and for $\theta_{23} =50^\circ$. For
$\theta_{23}=40^\circ$ the maximum is reached now at $\delta=\pm
180^\circ$ and not at $\delta=0^\circ$. For the $\theta_{23}=50^\circ$
case, the sensitivity curve bends in the opposite direction, due to
the opposite sign of $\cos 2 \theta_{23}$.

\begin{figure*} [ht]
\includegraphics[height=3in]{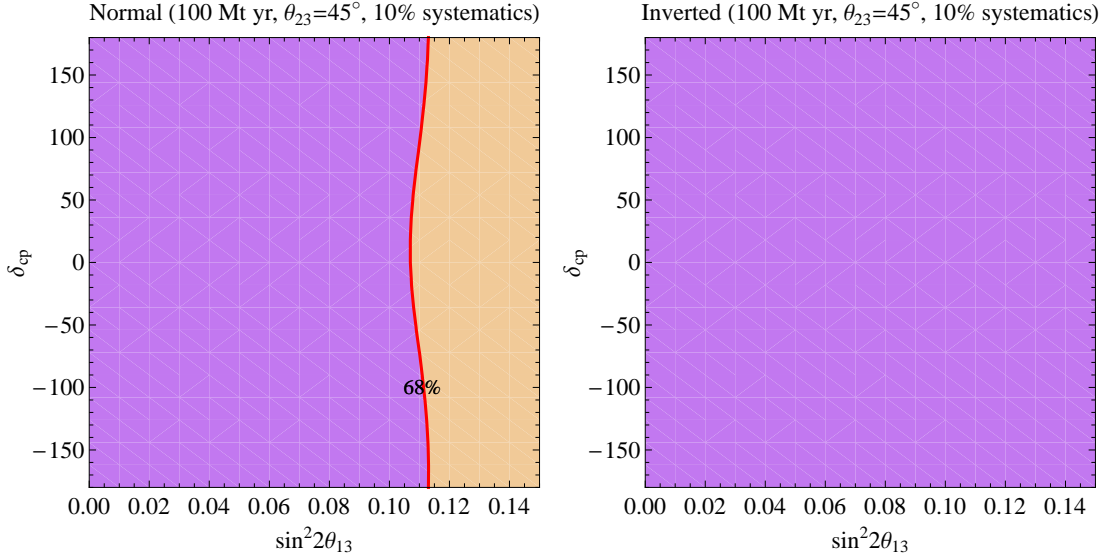}
\caption{ Left (right) pannel: hierarchy resolution using the
  muon-like contained events {in a detector with $100$~Mt~yr
  exposure,} for normal (inverted) hierarchies,
  atmospheric mixing angle $\theta_{23}=45^\circ$ as in
  Fig.~\ref{fig:fig6} but now marginalized over $|\Delta m^2_{31}|$.
  An overall $10\%$ systematic error has been included in the $\chi^2$
  analysis. }
\label{fig:dm31}
\end{figure*}

In Figure \ref{fig:dm31} we show the results for the case as in
Fig.~\ref{fig:fig6}, but here we also marginalize over the value of
$|\Delta m^2_{31}|$ in the allowed range.  The ability to distinguish
the hierarchy appears significantly decreased, after using the same 2
d.o.f statistics to define CL.  It is however important to note that
we expect the value of $|\Delta m^2_{31}|$ to be known with much
higher precision on the time scale relevamt for the IceCube deep core
experiments.

\section{Discussion and Conclusions}
\label{sec:disc}

The IceCube detector and its proposed deep core array provide a great
opportunity for studies of atmospheric neutrinos. Being the largest
existing neutrino detector, it will accumulate a huge number of
atmospheric neutrino events over an enormous energy range, thus
allowing for detailed studies of oscillation physics, Earth density,
atmospheric neutrino fluxes and new physics. In order to extract all
this information it is necessary to use energy and angular
distribution information, as well as flavour composition, all possible
to obtain with the IceCube detector.

Qualitatively, there are three main energy intervals and three main
angular regions which are sensitive to different types of physics.

At very high energies, above 10 TeV, neutrino interaction
cross-sections become high enough that neutrinos going through the
Earth start getting attenuated. This effect is sensitive to neutrino
interaction cross-sections and to the density profile of the Earth. A
measurement of the neutrino flux at these energies can provide a
determination of the Earth density \cite{density} which can be used as
an experimental input to our analysis, instead of the PREM
predictions.

The ``intermediate'' energy region, between 50 GeV and 1 TeV can
provide good information about the atmospheric neutrino flux, which
can be used to improve the uncertainties in the simulated atmospheric
neutrino fluxes.

In our paper we concentrated on the ``low'' energy region, below about
40 GeV, where neutrino oscillation effects can be significant. Matter
effects inside the Earth are very important in this energy range and
for non-zero values of $\theta_{13}$ resonance effects can strongly
enhance/reduce oscillation probabilities.  Straight-up-going neutrinos
($c_\theta\le -0.7$) pass through the core of the Earth and are most
sensitive to resonant matter oscillations and thus to all sub-dominant
neutrino oscillation effects ($\theta_{13}$, mass hierarchy, CP
violation).  Up-going neutrinos at shallower angles are still
sensitive to the ``main'' oscillation effects ($\Delta m^2_{31}$,
$\theta_{23}$), while the other, sub-dominant contributions become
smaller, due to the lower matter densities and shorter pathlengths.

In the present study we have exploited muon neutrinos and
antineutrinos with energies in the $10-30$~GeV range which have
crossed the Earth (i.e. $c_\theta\le -0.7$). Using the $\mu$ like
contained events for an exposure of 100 Mton yr in the proposed
Icecube deep core ice Cherenkov detector, the neutrino mass hierarchy
could be extracted at the $90\%$ CL if $\sin^2 2 \theta_{13}>0.02$
($\sin^2 2 \theta_{13}>0.14$ when a $10\%$ systematic error is
included in the analysis) regardless of the value of the CP violating
phase $\delta$.  In addition, downgoing neutrinos in the deep core
will provide a measurement of the atmospheric neutrino flux, helping
enormously in diminishing the systematic uncertainties \cite{atmflux}.

The Icecube deep core array, with muon energy detection threshold of
$\sim 5$~GeV could provide the first measurement of the neutrino mass
hierarchy if $\sin^2 2 \theta_{13}$ is close to the present upper
limit. This would be possible with an exposure of $100$~Mt~yr and if
systematic errors below 10\% can be achieved.  The next generation of
long baseline experiments would then have to concentrate only on the
extraction of the CP violating phase $\delta$.

\section*{Acknowledgments} 

We would like to thank Doug Cowen, Carsten Rott and especially Ty
DeYoung for useful discussions and comments.  We would also like to
thank Michele Maltoni for useful comments and checks of our
results. OM would like to thank the precious help provided by
E. Fern\'andez Mart\'{\i}nez for the numerical analysis.  This work
was supported in part by the NSF grant PHY-0555368, by the European
Programme ``The Quest for Unification'' contract MRTN-CT-2004-503369
and by a \emph{Ram\'on y Cajal} contract from MEC, Spain.  SR is a
National Research Council Research Associate at the Naval Research
Laboratory. I.M. would like to thank the UT Austin theory group for
hospitality while part of this work was being completed. This material
is based upon work supported in part by the National Science
Foundation under Grant No. PHY-0455649.

\end{document}